# Visibility Without Guidance: Measuring the Actionability of Public Crisis Communication in Lebanon

**Mohamed Soufan**
Independent Researcher & Data Scientist
Antalya, Turkey
ORCID: https://orcid.org/0009-0004-1705-5574
Email: m@soufan.tech

## Abstract

Public crisis communication is most valuable when it enables protective action: when it tells affected populations not only what is happening, but where, what to do, how to access help, and whom the guidance applies to. This study examines whether institutional crisis communication in Lebanon during an active conflict escalation met that standard. It presents a systematic actionability audit of 182 public-facing communication items issued by nine Lebanese state institutions, emergency responders, and United Nations humanitarian agencies between April 1 and May 7, 2026.

Each item was coded on five binary dimensions: specific location, clear instruction, access channel or contact, time or date reference, and target group identification. These dimensions were derived from established warning communication frameworks. An actionability score between 0 and 5 was calculated as the arithmetic sum of the five dimensions, and score consistency was mechanically verified across all items before analysis.

Findings indicate a substantial gap between communicative output and actionable guidance. The mean actionability score across all 182 items was 1.98 out of 5, with 46.7% of items scoring 0 or 1. The most consistently absent element was access channel or contact information, missing from 89.6% of items. Institutional variation was marked: UN humanitarian agencies produced a mean score of 3.10, compared with 1.35 for state and ministry sources, while emergency responders occupied an intermediate position. Communication type analysis shows that the prevalence of informational reporting over operational guidance helps explain the low aggregate score.

These results suggest that the dominant mode of institutional communication during this period was informative rather than directive: characterized by situational description rather than actionable specification. The findings contribute to the empirical literature on crisis communication adequacy in conflict-affected settings and offer a replicable audit framework applicable to comparable institutional contexts.

## Introduction

When populations face acute danger, the value of official communication is not measured by its volume. It is measured by whether a person who receives it knows what to do next. A post that reports airstrikes in south Lebanon tells its reader that something has happened. A post that tells them where the nearest shelter is, how to register for assistance, and which number to call tells them what to do about it. These are not subtle differences. In a crisis environment defined by displacement, infrastructure failure, and degraded access to services, the gap between informational visibility and actionable guidance is the gap between communication that functions and communication that does not.

This paper examines that gap empirically. We present a systematic audit of public-facing crisis communication issued by nine Lebanese state institutions, emergency responders, and United Nations humanitarian agencies during an active conflict escalation between April 1 and May 7, 2026. The period was defined by sustained Israeli military operations across south Lebanon, Beirut's southern suburbs, and the Bekaa Valley; displacement exceeding one million people; and severe pressure on health, shelter, and water systems (UNHCR, 2026; OCHA, 2026a). All nine institutions issued public communications during this period, though the volume and format varied substantially across sources. The question this study addresses is not whether they communicated, but whether what they communicated provided practical guidance to the populations they serve.

The broader literature on crisis communication has long identified actionability — the degree to which a communication enables a recipient to take a specific protective action — as a core dimension of warning system effectiveness. Sorensen defines effective warning systems as systems that detect impending danger, communicate that information to people at risk, and enable those in danger to make decisions and take action (Sorensen, 2000). Lindell and Perry's Protective Action Decision Model elaborates the cognitive conditions under which individuals convert received warnings into protective behaviour, identifying the processing of warning messages, protective-action options, and situational cues as central to response (Lindell and Perry, 2012). Empirical work in emergency warning communication has also shown that the content and clarity of warning messages affect whether warnings are understood, believed, and acted upon (Mileti and Sorensen, 1990). Yet systematic, item-level audits of institutional communication during active armed conflict — as opposed to natural disaster or public health emergency settings — remain comparatively rare, and studies applying a uniform quantitative framework across multiple institutional types simultaneously are rarer still.

The contribution of this study is threefold. First, we operationalise actionability as a five-dimensional binary coding framework — specific location, clear instruction, access channel or contact, time or date reference, and target group identification — and apply it uniformly across 182 communication items, yielding a dataset that is internally consistent and mechanically verified. Second, we apply this framework across an important selected cross-section of public-facing crisis communication in Lebanon, enabling comparison between individual sources and institutional categories: state ministries, emergency responders, and UN humanitarian agencies. Third, we analyse communication from a recent active phase of conflict escalation, offering a time-bounded snapshot of communication adequacy under conditions of high institutional demand.

The results show a substantial actionability gap. The mean actionability score across all 182 items is 1.98 out of 5. Nearly half of all items — 46.7% — score 0 or 1, meaning they provide at most one practically navigable element. Access channel or contact information, a critical element for populations seeking help, is absent from 89.6% of items. Institutional variation is substantial: UN humanitarian agencies produce a mean score of 3.10, compared to 1.35 for state and ministry sources. Emergency responders occupy an intermediate position but remain below the midpoint of the scale. These patterns are reinforced by communication-type analysis, which shows that situation reports and incident updates dominate higher-volume sources.

The paper proceeds by describing the data collection, coding framework, analytical approach, results, interpretation, limitations, ethics, and conclusion.

## Methodology

### Study Design and Research Framing

This study employs a systematic communication audit to assess the actionability of public-facing crisis communication issued by Lebanese state institutions, emergency responders, and United Nations humanitarian agencies during an active conflict escalation. The central question is not whether these actors communicated during the crisis — evidence that they did is abundant — but whether what they communicated was of practical use to affected populations. We operationalise this distinction through a five-dimensional actionability framework applied uniformly across all collected items, yielding a quantitative measure of the gap between informational visibility and functional guidance.

The framework draws on warning communication literature, including Sorensen's work on hazard warning systems and Lindell and Perry's Protective Action Decision Model, which together identify warning dissemination, message processing, protective-action options, and response capability as central to protective action (Sorensen, 2000; Lindell and Perry, 2012).

### Collection Window and Contextual Scope

The collection window spans April 1 to May 7, 2026 — a period defined by the active escalation of hostilities in Lebanon following Israeli military operations that intensified from early March 2026 (OHCHR, 2026). The window opens after the initial displacement shock had registered across relevant institutions and closes at the point of data collection, capturing a sustained phase of active emergency rather than onset or recovery. Selecting this bounded window allowed us to capture communication behaviour during a period of high institutional demand: when populations were displaced, infrastructure was degraded, and actionable information carried substantial practical value.

The conflict context during this period included continued Israeli airstrikes across south Lebanon, Beirut's southern suburbs, and the Bekaa; displacement of over one million people; damage to civilian infrastructure, including health facilities; and a ceasefire beginning on April 17, 2026, after which violations continued to be reported (OHCHR, 2026; OCHA, 2026b; Associated Press, 2026). This context is material to interpretation: the gap between visibility and guidance identified in this study cannot be attributed to a stable low-intensity environment in which guidance was simply unnecessary.

## Source Selection

Sources were selected to span an important cross-section of public-facing crisis communication in Lebanon: state ministries responsible for population welfare, dedicated emergency response bodies, and United Nations humanitarian agencies operating in-country. Nine sources were included across three institutional categories.

**State and ministry sources** (n = 3): the Disaster Risk Management Unit (DRM Lebanon), the Ministry of Social Affairs, and the Ministry of Public Health. These bodies hold central public responsibilities for emergency coordination, social assistance, and public health communication during crisis periods.

**Emergency responders** (n = 3): the Lebanese Civil Defense, the Lebanese Army, and the Lebanese Red Cross. Their communication is operationally relevant because these actors are directly involved in crisis response, rescue, casualty transport, UXO risk communication, and security-related public messaging.

**UN humanitarian agencies** (n = 3): UNHCR Lebanon, UNICEF Lebanon, and WFP Lebanon. These organisations operate under the Lebanon Response Plan and maintain dedicated public information functions. Their flash updates and press communications serve both coordination and public-facing purposes, and in several cases represent the most structured periodic reporting available during the window.

Sources were restricted to those maintaining an active, publicly accessible communication presence during the collection window. Where an organisation operated across multiple platforms, the primary platform was selected based on activity level and data accessibility: Facebook and/or official websites for Lebanese state institutions, X/Twitter for the Lebanese Army and Red Cross, flash update PDFs for UNHCR, and combined website, PDF, and X/Twitter sources for UNICEF and WFP. Secondary platforms were not included where they duplicated primary content.

**Table 1. Dataset composition by source**

| **Source** | **Category** | **Platform / Data type** | **Items** |
|---|---|---|---|
| DRM Lebanon | State/ministry | Facebook | 30 |
| Ministry of Social Affairs | State/ministry | Facebook | 5 |
| Ministry of Public Health | State/ministry | Facebook / Website | 42 |
| Lebanese Civil Defense | Emergency responder | Facebook | 27 |
| Lebanese Army | Emergency responder | X (formerly Twitter) | 40 |
| Lebanese Red Cross | Emergency responder | X (formerly Twitter) | 7 |
| UNHCR Lebanon | Humanitarian/UN | Flash updates (PDF) | 4 |
| UNICEF Lebanon | Humanitarian/UN | Press releases / reports | 6 |
| WFP Lebanon | Humanitarian/UN | X / website | 21 |
| **Total** | | | **182** |

*Items are original public-facing institutional outputs published April 1 – May 7, 2026. Retweets, quoted posts, and non-crisis content were excluded prior to coding.*

## Data Collection

Data collection was conducted through platform-specific methods suited to each source's publishing format. For Facebook-based sources (DRM Lebanon, Ministry of Social Affairs, Ministry of Public Health, Lebanese Civil Defense), posts were collected via structured web scraping of public page timelines within the collection window. For X-based sources (Lebanese Army, Lebanese Red Cross, WFP Lebanon), posts were extracted from source-specific JSON exports filtered to the collection window using ISO timestamp parsing, excluding retweets and quoted content to retain only original institutional communication. For document-based and website-based sources, publicly available flash updates, reports, and press releases were downloaded or extracted for analysis.

All raw data was preserved in source-specific directories prior to cleaning. Cleaning operations included removal of platform UI artefacts, exclusion of non-crisis content (ceremonial posts, donor announcements unrelated to operational response, generic institutional slogans), and exclusion of duplicate items identified by URL or normalised title matching. Public comments were excluded in all cases; only original institutional output was retained.

The resulting cleaned dataset contains 182 items across nine sources, covering the April 1 – May 7, 2026 window, with no historical leakage identified through systematic timestamp verification.

## Unit of Analysis

The unit of analysis is the individual public-facing communication item: a single post, press release, or periodic report issued by an institution within the collection window. For social media sources, each post constitutes one item. For document-based sources (UNHCR and UNICEF flash updates), each document constitutes one item, reflecting that these reports are issued as unified public communications on a weekly cycle. This treatment preserves the natural granularity of institutional communication behaviour: a single flash update represents one act of public communication regardless of its internal length or sectoral coverage.

Items were retained only where they met three conditions: originating from the identified institutional account or channel; published within the collection window; and classifiable as crisis-related based on content. Non-crisis items were excluded and documented in source-level notes.

## Actionability Coding Framework

Each item was coded on five binary dimensions derived from the warning communication literature and the practical requirements of civilian protective action:

| Dimension | Field | Positive condition |
|---|---|---|
| Specific location | location | Item names a specific geographic area relevant to the crisis |
| Clear instruction or action | instruction | Item tells recipients what to do or not to do |
| Access channel or contact | contact_link | Item provides a hotline, website, physical address, or referral pathway |
| Time or date reference | `time_date` | Item includes a specific time, date, or temporal marker |
| Target group identified | `target_group` | Item addresses a defined population segment |

Each dimension was scored as binary (`yes`/`no`). The actionability score for each item is the arithmetic sum of the five binary fields, yielding an integer between 0 and 5. A score of 0 indicates a communication item that conveys information but provides no practically navigable content; a score of 5 indicates an item that specifies where something is happening, what affected people should do, how they can access help, when this applies, and to whom it applies.

This scoring logic is intentionally conservative. An item that names a location but provides no instruction scores 1; naming a location and issuing an instruction without providing a contact channel scores 2. The framework does not reward frequency of output, emotional salience, or institutional prestige — only the presence of elements that enable a recipient to take a specific protective action.

All five binary fields and the derived score were verified mechanically: the recorded `actionability_score` for every item was confirmed to equal the sum of its five binary fields, with no manual overrides permitted. Items where initial manual coding produced inconsistencies were corrected to match the binary sum. This verification was applied across all 182 items prior to analysis.

## Coding Consistency and Validation

Coding was performed by the primary researcher with reference to explicit decision rules for each binary dimension. The instruction field, which carries the highest interpretive burden, was coded positively only where the item explicitly directed recipients toward a specific behaviour (e.g., "do not approach unexploded ordnance," "ensure your child's vaccinations are up to date," "follow army guidance before returning south"). Implicit or normative statements ("the situation remains difficult," "UNICEF calls on all parties") were coded negatively. The contact_link field was coded positively only where a functional access mechanism was named — a WhatsApp number, hotline, website, or named physical facility — not where a general organisational presence was described. Mechanical verification of score consistency across all 182 items served as a systematic validation check on coding coherence.

## Source-Type Classification

For aggregate comparison, sources were classified into three institutional categories: state/ministry (DRM Lebanon, Ministry of Social Affairs, Ministry of Public Health), emergency responder (Lebanese Civil Defense, Lebanese Army, Lebanese Red Cross), and humanitarian/UN (UNHCR Lebanon, UNICEF Lebanon, WFP Lebanon). This classification reflects both organisational mandate and the structural position each type of actor occupies within the crisis response architecture. Differences in actionability across these categories are interpreted as indicators of broader communication patterns, while avoiding claims about institutional intent.

## Analytical Approach

Five analyses were conducted on the cleaned and verified dataset. First, overall actionability was characterised through the mean score and the distributional breakdown across low (0–1), medium (2–3), and high (4–5) bands. Second, the prevalence of each binary dimension was calculated across all 182 items to identify which elements of actionable communication were most consistently absent. Third, items were grouped by communication type and mean actionability scores were calculated per type, identifying whether the dominance of particular formats explains the low aggregate scores. Fourth, sources were compared individually on item count, mean score, percentage of high-actionability items, and dominant communication type. Fifth, the three source-type categories were compared on the same metrics to support an institutional-level interpretation.

No inferential statistics were applied because the study analyses the cleaned set of crisis-relevant items collected from the selected source-platforms during the study window. The goal is descriptive auditing rather than population inference.

## Results

Overall Actionability

Across all 182 items, the mean actionability score is 1.98 out of 5. The score distribution is skewed toward the lower end of the scale. Eighty-five items (46.7%) score 0 or 1, indicating that these communications satisfy at most one of the five actionability criteria. Seventy-one items (39.0%) fall in the middle band, scoring 2 or 3. Twenty-six items (14.3%) score 4 or 5, representing communications that satisfy most or all of the framework criteria. High-actionability items constitute a minority of total institutional output during the study window.

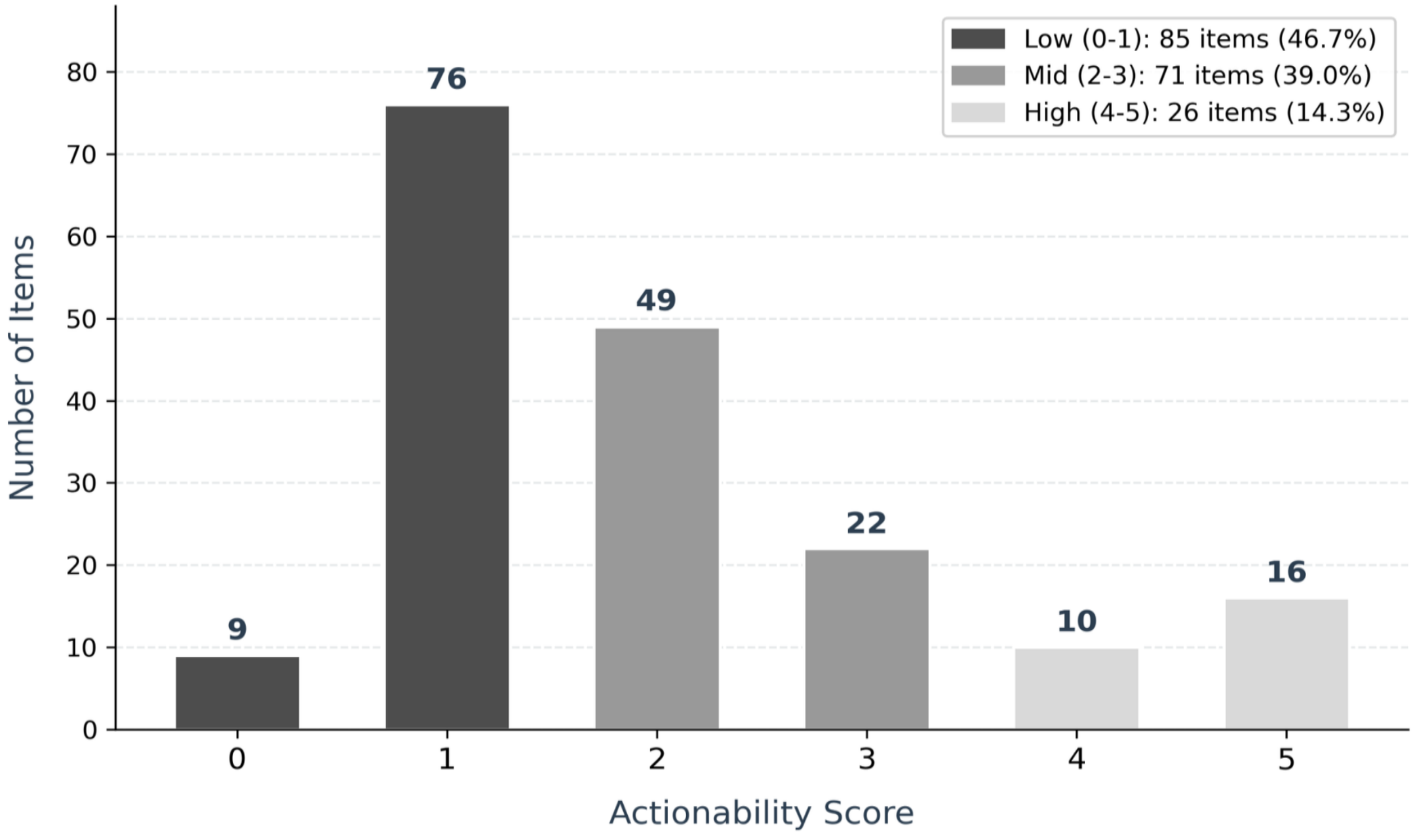

Figure 1. Distribution of actionability scores across 182 public-facing crisis communication items. Scores range from 0 to 5, corresponding to the number of actionability elements present in each item. Low scores (0–1) account for 46.7% of items, mid-range scores (2–3) for 39.0%, and high scores (4–5) for 14.3%.

## Prevalence of Actionability Elements

The five binary dimensions vary substantially in their prevalence across the dataset. Time or date reference is the most consistently present element, appearing in 147 items (80.8%). Specific location is present in 108 items (59.3%). The three remaining dimensions are each present in fewer than one in three items. Target group identification is present in 51 items (28.0%). Clear instruction or action is present in 35 items (19.2%). Access channel or contact information is the least represented dimension across the dataset, present in only 19 items (10.4%); it is absent from 163 items, representing 89.6% of the total corpus.

## Communication Type and Actionability

The dataset spans nineteen communication type categories. Three types account for 72.0% of all items: situation reports (n = 75, 41.2%), incident reports (n = 32, 17.6%), and coordination updates (n = 24, 13.2%). These dominant types are associated with below-average actionability scores. Situation reports yield a mean score of 1.57, with 10.7% of items in the high band. Coordination updates yield a mean score of 1.25, with 4.2% reaching the high band. Incident reports yield a mean score of 2.00, with no items in the high band.

Less frequent communication types are associated with markedly higher scores. Warnings and alerts (n = 14) yield a mean score of 3.64, with 57.1% of items scoring in the high band. Service announcements (n = 7) yield a mean score of 3.29, with 57.1% scoring in the high band. The two hotline or contact guidance items each score 5.00. Political and ceremonial items (n = 3) yield the lowest mean score in the dataset at 0.67.

## Source-Level Comparison

Actionability scores vary considerably across the nine sources. UNHCR Lebanon and UNICEF Lebanon produce the highest mean scores in the dataset, at 4.75 (n = 4) and 4.50 (n = 6) respectively; all UNHCR items and 83.3% of UNICEF items fall in the high band, and neither source records any item in the low band. WFP Lebanon, the third UN agency, produces a mean score of 2.38 (n = 21), with all items falling in the mid-range band and none in either the high or low band. The Lebanese Army is the highest-scoring emergency responder, with a mean score of 2.80 (n = 40) and 30.0% of items in the high band.

The remaining five sources produce mean scores below 1.50. The Ministry of Public Health yields a mean of 1.50 (n = 42), with 69.0% of items in the low band. Lebanese Civil Defense yields a mean of 1.44 (n = 27), with 63.0% in the low band. The Lebanese Red Cross yields a mean of 1.29 (n = 7), with no items in the high band. DRM Lebanon yields a mean of 1.20 (n = 30), with 90.0% of items in the low band — the highest concentration of low-scoring items across any single source. The Ministry of Social Affairs yields the lowest mean score at 1.00 (n = 5), with 60.0% of items in the low band.

## Source-Type Comparison

Aggregating by institutional category, UN humanitarian agencies produce the highest mean actionability score (3.10, n = 31), with 29.0% of items in the high band and no items in the low band. Emergency responders produce an intermediate mean score (2.16, n = 74), with 18.9% of items in the high band and 35.1% in the low band. State and ministry sources produce the lowest mean score (1.35, n = 77), with 3.9% of items in the high band and 76.6% in the low band. State and ministry sources account for 42.3% of all items in the dataset, making them the largest contributor by volume and the lowest by mean actionability score.

## Discussion

The findings of this audit point to a consistent and measurable gap between institutional communication activity and actionable guidance during the study period. Across 182 items drawn from nine institutions, the mean actionability score of 1.98 out of 5 indicates that the typical public communication issued during this crisis period satisfied fewer than two of the five criteria necessary for a recipient to take a specific protective action. This result is not a product of silence: all nine institutions communicated, and several did so at high volume. The gap identified is not one of presence but of content.

It is important to note at the outset that this framework evaluates public-facing communication outputs only. The findings speak to what was published and made accessible to the public during the study window; they do not constitute an assessment of the operational effectiveness, internal coordination capacity, or institutional response of the organisations included.

The dominance of the informational register

The communication-type analysis offers a structural explanation for the aggregate scores. Situation reports, incident reports, and coordination updates together account for 72.0% of all items in the dataset, and all three types produce below-average actionability scores. These formats are characterised by the reporting of conditions — locations of incidents, casualty figures, infrastructure status, institutional activities — rather than the issuance of guidance. They answer the question of what is happening rather than what affected populations should do. The high prevalence of time and date references (80.8% of items) alongside low prevalence of instructions (19.2%) and contact information (10.4%) is consistent with this pattern: institutions are timestamping their outputs and locating events geographically, but are not consistently translating that situational awareness into navigable guidance for recipients.

This pattern aligns with warning communication research showing that effective warnings require more than hazard notification: they must also communicate specific protective-action guidance and support public response (Mileti and Sorensen, 1990; Sorensen, 2000). Lindell and Perry's Protective Action Decision Model further specifies that protective response depends not only on information about a threat, but also on perceptions of available protective actions and the feasibility of taking them (Lindell and Perry, 2012). The near-universal absence of contact or access channel information — present in only 10.4% of items — represents a substantial gap at this stage of the warning process: populations may receive accurate situational information and yet have no pathway to act on it.

## Institutional variation and structural capacity

The divergence between source types is substantial and warrants consideration as a structural rather than incidental finding. UN humanitarian agencies produce a mean score of 3.10, compared to 1.35 for state and ministry sources — a difference of 1.75 points on a five-point scale. This divergence is not explained by volume alone: UN agencies contribute only 17.0% of items in the dataset, yet consistently produce the highest-scoring communications. UNHCR Lebanon and UNICEF Lebanon, both operating on periodic flash update formats, produce mean scores of 4.75 and 4.50 respectively, with no items in the low band.

The flash update format is relevant here. Structured periodic reports issued on a regular cycle and designed for both coordination and public communication are architecturally predisposed toward actionability: they include date references, geographic scope, target population identification, and operational guidance as standard components. The high scores observed for these sources may therefore reflect format design as much as communication intent. This observation is consistent with the finding that WFP Lebanon, whose output is dominated by operational updates rather than flash reports, produces a mean score of 2.38 despite belonging to the same institutional category — suggesting that format is a significant mediating variable.

State and ministry sources present a contrasting picture. With 42.3% of all items in the dataset and a mean score of 1.35, these sources are the most active communicators by volume while producing the lowest actionability scores across the institutional categories examined. DRM Lebanon — the institution nominally responsible for disaster risk coordination — records 90.0% of its items in the low band, the highest such concentration in the dataset. The Ministry of Social Affairs and the Ministry of Public Health follow a similar pattern, with 60.0% and 69.0% of their items respectively scoring 0 or 1.

Emergency responders occupy an intermediate position, though with considerable internal variation. The Lebanese Army produces a mean score of 2.80 with 30.0% of items in the high band, reflecting a communication style that more frequently includes location specificity and operational instruction. The Lebanese Civil Defense and Lebanese Red Cross produce mean scores below 1.50, with neither source recording any items in the high band. The Civil Defense in particular, whose operational role involves direct public-facing intervention, records 63.0% of items in the low band.

## The contact information deficit

The limited presence of access channel or contact information across the dataset — present in only 19 items (10.4%) — merits specific attention. In crisis conditions characterised by displacement and infrastructure disruption, the ability to identify a point of contact — a hotline, a registration platform, a named facility — is a precondition for accessing the services that institutions are simultaneously announcing. Communications that describe available services without providing a means of access satisfy the conditions of visibility without enabling the behaviour they implicitly invite. This gap is most pronounced among state and ministry sources but is not confined to them.

### Format asymmetry and interpretation

As noted in the limitations section of this paper, document-based sources such as UNHCR and UNICEF flash updates are structurally longer than social media posts and have more opportunity to satisfy multiple binary dimensions within a single item. The high scores observed for these sources are therefore not straightforwardly comparable to scores observed for social media posts, which are constrained by format. This asymmetry does not invalidate the findings — the actionability framework measures what is present, regardless of format — but it cautions against treating score differences as reflecting solely communication intent. A more granular comparison would require controlling for format length and post type, which falls outside the scope of the present study.

### Implications for practice

The findings suggest that actionability deficits in institutional crisis communication are not uniform but are concentrated in specific institutional categories and communication formats. Interventions aimed at improving communication adequacy during conflict emergencies would benefit from targeting the structural dimensions of the deficit: the adoption of standardised templates that embed actionability elements by design, the consistent inclusion of contact and access channel information across all communication types, and the reorientation of high-volume formats — particularly situation reports and coordination updates — toward directive as well as informational content. The divergence between format-driven and ad hoc communication outputs suggests that design rather than intent may be the more tractable lever for improvement.

## Limitations

Several limitations bear on the interpretation of findings. First, source coverage is selective: nine institutions were included based on public accessibility, institutional mandate, and relevance to crisis communication, rather than statistical representativeness; other actors active during the crisis — municipal authorities, political party channels, civil society organisations, and informal community networks — were outside the scope of the study. Second, the study is restricted to public-facing communication on the selected primary platform for each source; internal coordination messages, private group communications, and platforms not captured by the collection method are not reflected. Third, comments, replies, and citizen-generated content were excluded throughout; the dataset captures only institutional output. Fourth, coding was conducted by a single researcher. Although actionability scores were mechanically verified as the arithmetic sum of the five binary fields across all 182 items, the absence of a second independent coder limits assessment of interpretive reliability, particularly for the instruction and target group dimensions. Fifth, the actionability framework measures the presence of practical communication elements, not their real-world effectiveness: a post scoring 5 may still have failed to reach its intended audience, and a post scoring 1 may still have reached a large audience. Score and impact are analytically distinct. Sixth, document-based sources, particularly UNHCR and UNICEF flash updates, are structurally longer and more information-dense than social media posts; comparisons of actionability scores across source types should account for this format asymmetry, as longer documents have more opportunities to satisfy multiple binary dimensions within a single item.

## Ethics

This study analyses publicly available institutional communication. All items in the dataset were published by the identified organisations on publicly accessible platforms or official websites during the study window and were collected without circumvention of access restrictions of any kind. No personally identifiable information was collected, processed, or stored at any stage of the research. The dataset contains institutional outputs only; individual user content, comments, and replies were excluded throughout.
The study does not involve human participants, does not collect sensitive personal data, and does not engage with vulnerable individuals directly. Because the study analyses only public institutional outputs, does not involve interaction with individuals, and does not process private personal data, it would normally fall outside the scope of human-subjects ethics review.
Source institutions are identified by name in accordance with standard practice in public communication audits. Identification is necessary for the transparency and reproducibility of the findings and is consistent with the public nature of the communications examined. No claims are made about the internal operations, private conduct, or personnel of any institution included in the study.

## Conclusion

This study set out to determine whether public-facing crisis communication issued by Lebanese state institutions, emergency responders, and United Nations humanitarian agencies during an active conflict escalation was actionable — that is, whether it provided affected populations with the specific elements necessary to take a protective action. The findings indicate that, in the majority of cases, it did not.

Across 182 items collected between April 1 and May 7, 2026, the mean actionability score is 1.98 out of 5. Nearly half of all items provide at most one of the five actionability elements. Access channel and contact information — the element that connects situational awareness to practical response — is absent from 89.6% of the dataset. Low actionability is most pronounced among state and ministry institutions, which account for the largest share of total output.

The study also identifies substantial variation across institutional categories. UN humanitarian agencies, especially those operating through structured periodic reporting formats, produce higher actionability scores. State and ministry sources, producing the highest volume of communications, produce the lowest. This divergence suggests that the actionability of institutional crisis communication is shaped in significant part by format design and communication infrastructure rather than by volume of output alone.

Three contributions follow from these findings. First, the study provides an empirical baseline for assessing public-facing crisis communication adequacy in Lebanon during this period, against which future communication practices can be compared. Second, it demonstrates the viability of a uniform, mechanically verified actionability framework applied across heterogeneous institutional sources and communication types — a methodological approach transferable to comparable crisis contexts. Third, it adds to the limited body of quantitative communication audits conducted during active armed conflict, a setting that remains underrepresented relative to natural disaster and public health emergency contexts in the existing literature.

Future research could extend this framework in several directions: longitudinal application across multiple crisis phases to examine whether actionability changes with conflict intensity; cross-country comparison of institutional communication patterns in conflict-affected settings; and integration of audience reach and language accessibility data to move from measuring communicative output toward measuring communicative impact.

The gap between visibility and guidance documented here is not inevitable. It reflects, at least in part, structural and design choices that are amenable to intervention. Crisis communication that is seen but not navigable leaves populations informed of danger without the means to respond to it.

## Supplementary Materials

The following supplementary materials accompany this paper: (S1) the full coded dataset in CSV format, including all 182 items with source metadata, binary field codings, and derived actionability scores; (S2) the Python analysis script used to generate all quantitative outputs reported in the Results section; (S3) the coding decision rules applied to each binary dimension, including annotated examples of positive and negative coding decisions for the instruction and contact_link fields. All supplementary files are available at the repository listed in the Data Availability Statement.

## Data Availability Statement

The dataset and analysis scripts supporting the findings of this study are openly available on GitHub at: https://github.com/mohamedsoufan/lebanon-crisis-communication-actionability
The repository contains the full cleaned dataset, the five analysis output files, and the Python script used to produce the results. Raw source files are included where platform terms of service permit redistribution. Full-text content from Facebook and X/Twitter posts is not redistributed in the repository because platform policies may restrict republication of collected post content. For these items, the dataset provides source metadata, URLs where available, timestamps, classifications, binary coding fields, and derived actionability scores sufficient to reproduce the analysis from the cleaned coding layer. Items collected from other public sources — including UNHCR and UNICEF flash update documents — are included in full as publicly available institutional publications.

## Author Contributions

Mohamed Soufan conducted all aspects of the study, including conceptualization, methodology, software development, formal analysis, investigation, data curation, visualization, and writing.

Acknowledgments

The author acknowledges that the study relies on publicly available institutional communications issued by the organisations included in the dataset. The author also acknowledges the open-source tools and publicly accessible data services that supported data collection and analysis.

Use of Generative AI

During the preparation of this manuscript, the author used ChatGPT to assist with language polishing, structural editing, and clarity of presentation. The author reviewed and verified the final manuscript, analysis, citations, and all substantive claims, and remains solely responsible for the content.

Funding

This research received no external funding.

Conflicts of Interest

The author declares no conflicts of interest.